\documentstyle[12pt,epsf,axodraw]{article}
\def\gsim{\mathrel{\rlap {\raise.5ex\hbox{$ > $}}
{\lower.5ex\hbox{$\sim$}}}}
\def\lsim{\mathrel{\rlap {\raise.5ex\hbox{$ < $}}
{\lower.5ex\hbox{$\sim$}}}}

\newcommand{\pr}{\paragraph{}}
\newcommand{\be}{\begin{equation}}
\newcommand{\ee}{\end{equation}}
\newcommand{\bea}{\begin{eqnarray}}
\newcommand{\nn}{\nonumber}
\newcommand{\eea}{\end{eqnarray}}

\def\gappeq{\mathrel{\rlap {\raise.5ex\hbox{$>$}}
{\lower.5ex\hbox{$\sim$}}}}
 
\def\lappeq{\mathrel{\rlap{\raise.5ex\hbox{$<$}}
{\lower.5ex\hbox{$\sim$}}}}
 
\begin{document}
 
\begin{titlepage}
\begin{flushright}

%CERN-TH/99-105 \\
ACT-3/99   \\
CTP-TAMU-18/99 \\
OUTP--99--05P \\
gr-qc/9904068 \\
\end{flushright}

\begin{centering}
\vspace{.1in}

{\Large\bf{Quantum-Gravitational Diffusion and
Stochastic Fluctuations in the Velocity of Light}}\\[15mm]

\vspace{.1in}

{\bf John Ellis}$^{a}$, 
{\bf N.E. Mavromatos}$^{a,b,\diamond}$ 
and {\bf D.V. Nanopoulos}$^{c}$ \\

\vspace{.5in}
 
{\bf Abstract} \\
\vspace{.1in}
\end{centering}
{\small 
We argue that quantum-gravitational fluctuations in the
space-time background give the vacuum non-trivial
optical properties that include diffusion and
consequent uncertainties in the arrival times
of photons, causing stochastic fluctuations in the
velocity of light {\it in vacuo}. Our proposal is motivated within
a Liouville string formulation of quantum gravity
that also suggests a frequency-dependent refractive index
of the particle vacuum. We construct an explicit realization
by treating photon propagation through
quantum excitations of $D$-brane fluctuations
in the space-time foam. These are described by higher-genus
string effects, that lead to stochastic
fluctuations in couplings, and hence in the velocity of light.
We discuss the possibilities of
constraining or measuring photon diffusion {\it in vacuo} via
$\gamma$-ray observations of distant astrophysical sources.}

\vspace{1.in}
\begin{flushleft}
$^{a}$ CERN, Theory Division, Geneva CH-1211, Geneva 23, Switzerland.\\
$^{b}$ University of Oxford, 
Department of Physics, Theoretical Physics,
1 Keble Road,
Oxford OX1 3NP, U.K.  \\
$^{c}$ Department of Physics, 
Texas A \& M University, College Station, \\
TX~77843-4242, USA, \\
Astroparticle Physics Group, Houston
Advanced Research Center (HARC), Mitchell Campus,
Woodlands, TX 77381, USA, and \\
Academy of Athens, Chair of Theoretical Physics, 
Division of Natural Sciences, 28~Panepistimiou Avenue, 
Athens 10679, Greece. \\
$^{\diamond}$ P.P.A.R.C. Advanced Fellow.

\end{flushleft}

\end{titlepage} 

\section{Introduction} 

The propagation of light through media with
non-trivial optical properties is subject to three
important effects: one is a variation in the light
velocity with photon energy, namely a frequency-dependent refractive
index, a second is a difference between the velocities of light with
different polarizations, namely birefringence,
and the third is a diffusive spread in the apparent velocity of
light. We have argued previously that quantum-gravitational
fluctuations in the space-time background may endow the
conventional particle vacuum with such non-trivial optical
properties, in particular a frequency-dependent refractive
index~\cite{aemn}. We have also observed that such an effect may be
severely
constrained
by careful observations of distant astrophysical objects whose
emissions exhibit short time structures, such as Gamma-Ray Bursters
(GRBs)~\cite{nature}. The possibility of quantum-gravitational
birefringence has
been raised within a loop approach to quantum gravity~\cite{pullin}. The
purpose of this paper is to propose how quantum-gravitational
diffusion may spread the arrival times of photons from distant
sources, even if they have the same energies (frequencies), 
corresponding to stochastic
fluctuations in the apparent velocity of light.

An example of this phenomenon in a conventional
optical medium, which illustrates clearly
the intuition behind our proposal, has been 
discussed in~\cite{AMANDA}. Light propagating through
ice may encounter air bubbles, which have a different
refractive index and hence induce scattering and diffusion.
We have argued previously~\cite{aemn} that foamy fluctuations 
in space-time generate a quantum-gravitational
`aerogel' with an effective refractive index
$\sim E/M_{QG}$ {\it in vacuo}, where $M_{QG}$ is some
characteristic scale that may approach the Planck mass $M_P$.
Now we argue that the quantum uncertainties in these
foamy quantum-gravitational fluctuations in turn induce
fluctuations in the refractive index that act as scattering
centres analogous to the air bubbles of~\cite{AMANDA}.
As illustrations of these quantum-gravitational fluctuations,
one might consider foaming microscopic black holes that
induce local fluctuations in the wave front close to their
horizons, leading on larger distance scales to diffusive
broadening of any light pulse.

In this paper, we provide a mathematical formulation of
this physical intuition, using the framework for quantum gravity
adopted in~\cite{aemn}, namely that
of Liouville strings~\cite{emn}, in which classical
conformal string backgrounds are allowed to exhibit
quantum space-time fluctuations. These cause
departures from criticality and conformal symmetry
that may be counterbalanced by introducing on the string world sheet
a Liouville field with non-trivial dynamics. 
This Liouville field may be considered as a dynamical
renormalization scale, and conformal invariance
may be restored by dressing operators with appropriate
Liouville factors~\cite{ddk}. Since the quantum-gravitational space-time
fluctuations
drive the string super-critical, the Liouville field has
negative metric, and we identify it with target time~\cite{emn,kogan}.

Interactions of the Liouville field with the conventional
string degrees of freedom are controlled by the Zamolodchikov flow~\cite{zam}
of the world-sheet renormalization-group equations.
These incorporate the conventional Hamiltonian structure
and $S$-matrix description of the scattering of
low-energy particles, but also encode information 
on the energy-dependent interactions of the probe with
non-local quantum-gravitational degrees of freedom. 
These fail to decouple~\cite{emn} 
from the low-energy matter in the presence 
of foamy singular metric fluctuations. Since
a low-energy observer cannot detect
such global modes by direct scattering experiments,
integrating them out of the low-energy effective
theory makes
the low-energy system resemble
an `open' quantum system interacting with an 
unobserved `environment' of these 
quantum-gravitational solitonic states. 

The outline of this paper is as follows. In Section 2, we
review relevant features of this Liouville string
framework~\cite{emn} for space-time foam,
discussing in particular the formal basis for the
appearance of stochastic fluctuations in theory space. 
Next, in section 3, we construct a
specific realization of this non-critical string
approach by considering quantum fluctuations
of $D$-brane excitations in the vacuum. Then, in Section 4,
we discuss the propagation of photons through this quantum
$D$-brane foam. Section 5 shows how stochastic
fluctuations occur in the context of $D$ branes,
and exhibits the expected diffusion
effect~\footnote{We do not find any evidence to support the
suggestion of birefringence~\cite{pullin}.}. Finally, in
Section 6, we discuss how this
phenomenon may be constrained or detected by observations
of distant astrophysical objects.

\section{Relevant Aspects of Liouville String}

In this Section we review briefly features of
Liouville that play roles in our subsequent discussion.
In this approach, one treats quantum fluctuations in the
space-time background as deviations from a classical
string background that is described by supercritical
deformations of a conformal field theory
on the world sheet. One may
restore criticality by Liouville dressing of
world-sheet model fields~\cite{ddk}. Specifically, 
consider a conformal $\sigma$ model, described by an action $S^*$ on the 
world sheet $\Sigma$, subject to non-conformal deformations 
$\int_{\Sigma}g^iV_id^2\sigma$, with $V_i$ appropriate vertex operators:
\be     
S_g = S^* + \int_{\Sigma}g^iV_id^2\sigma
\label{sigma}
\ee
The non-conformal nature of the couplings $g^i$ implies that their 
world-sheet renormal- \\
ization-group $\beta$ functions $\beta^i$
do not vanish. The following is the
generic structure of such $\beta$ functions, close 
to such a fixed point $\{g^{i*} = 0\}$:
\be
\beta^i = (h_i - 2)g^i + c^i_{jk}g^jg^k + o(g^3).
\label{fixed}
\ee
One cancels the deviation from criticality by world-sheet
`gravitational' dressing that corresponds to defining renormalized
couplings in a curved space. To ${\cal O}(g^2)$, one
has~\cite{ddk,schmid,dorn}:
\be
\lambda^i(t) = \lambda^ie^{\alpha_it} + \frac{\pi}{Q \pm 
2\alpha_i}c^i_{jk}\lambda^j\lambda^kte^{\alpha_it} + O(\lambda^3), \qquad
Q^2 = c-25 \nonumber
\label{renorm}
\ee
where $t$ is the zero mode of the Liouville field, $Q^2$ is the central 
charge deficit which is $\ge 0$ for the supercritical string case
of interest here, and the
$\alpha_i$ are gravitational anomalous
dimensions:
\be
\alpha_i(\alpha_i + Q) = h_i - 2 \qquad  
{\rm for} \qquad  c \ge 25 \nonumber
\label{anom}
\ee
The supercriticality implies
a Minkowskian signature for the Liouville field~\cite{aben}.
enabling us to identify its zero mode with target
time~\cite{emn,kogan}~\footnote{
We note for later use that~\cite{dorn}, in the
gravitationally-dressed (Liouville) world-sheet theory, 
only the leading-order coefficients in the $\beta$ functions 
are renormalization-scheme independent. This implies that,
when one identifies the Liouville field with target time, 
one loses general covariance in the foamy ground state.
This should be thought of as a spontaneous breaking 
of the symmetry by quantum-gravitational fluctuations.}. 
After the renormalization (\ref{renorm}), the critical-string conformal 
invariance conditions corresponding to the vanishing of flat-space 
$\beta$ functions are replaced by~\cite{ddk,schmid,dorn,emn}:
\be
{\ddot \lambda}^i + Q{\dot \lambda}^i = -\beta^i  \qquad 
{\rm for} c \ge 25. \nonumber
\label{neweq}
\ee
The minus sign in front of the flat-world-sheet $\beta$ functions
reflects the supercriticality of the string. 
\pr
The propagation of non-relativistic light-particle modes was
examined in~\cite{emn}, where a modification of the
quantum Liouville equation for the density matrix was found,
as proposed in~\cite{ehns}. The propagation of
massless probes was examined in~\cite{aemn},
where it was found that the conventional relativistic
energy-momentum dispersion relation is modified in
non-critical 
Liouville strings, as a result of the interaction with 
the quantum-gravitational environment,
as we now review briefly for the benefit of non-expert readers.
\pr
The first step is to observe that, in the case of interest, the
non-criticality of the $\sigma$ model describing the 
effective theory is induced by 
the operator-product-expansion coefficients 
$c^i_{j{\hat k}}$ that express the interaction of a low-energy 
probe (latin indices) with 
quantum-gravitational modes (hatted latin indices).
Hence, $Q^2 =0$ to lowest order, and the 
Liouville anomalous-dimension coefficients $\alpha _i$ 
are given simply by the magnitude of the spatial 
momentum of the massless probe: 
\be
    \alpha _i = |{\overline k}|
\label{spatial}
\ee
We use this to rewrite
(\ref{renorm}) approximately, to order $g^2$, as~\cite{aemn}: 
\be
     \lambda ^i(t) \simeq g^i e^{(\alpha _i + \Delta \alpha _i )t}
\label{shift2}
\ee
where the shift $\Delta \alpha _i$ is given by:
\be
      \Delta \alpha _i \simeq \frac{\pi}{2\alpha _i}c^i_{j{\hat k}}g^{{\hat k}}
\label{shift}
\ee
We next make the generic hypothesis, which is supported
by some explicit examples~\cite{kanti}, especially in the context of $D$ 
branes~\cite{dbranes}, as discussed in more detail below,
that
\be
\frac{1}{\alpha _i}c^i_{j{\hat k}} \sim \xi E/M_{QG},
\label{estimate} 
\ee
where $E$ is the energy scale of the low-energy probe, 
$\xi =\pm {\cal O}(1)$ and $M_{QG}$
is a characteristic 
of gravitational interactions, possibly to be identified
with the Planck scale $M_P \sim 10^{19}$ GeV. We infer from
(\ref{spatial}), (\ref{shift2}), (\ref{shift}) and
(\ref{estimate}) a modified dispersion relation
\be
E = |{\overline k}| \times \left( 1 + \xi E/M_{QG} \right)
\label{dispersion}
\ee
corresponding to an energy- (frequency-)dependent
refractive index $\eta = 1 + {\cal O}(E / M_{QG})$.
\pr
It was observed in~\cite{nature} that such a
quantum-gravitational effect
could be probed by observations of
distant astrophysical sources, such as GRBs, 
Active Galactic Nuclei (AGNs) or pulsars,
and that these could place important constraints
on models in which such a refractive index is suppressed 
only by a single power of $1/M_{QG}$, potentially with sensitivity
to $M_{QG} \sim M_P$. Here we go further, remarking that 
quantum effects due to world-sheet topology
fluctuations~\cite{emnquantum},
which arise in the context of first-quantized strings
from the summation over genera, may cause
observable stochastic effects that lead to a diffusive spread in
apparent velocities even for photons of fixed energy
(frequency). This would imply the spreading of an
initial pulse, and place limitations  
on the resolutions of experimental 
measurements of distant astrophysical sources.
\pr
The basic motivation for this suggestion comes from the
quantization of the target-space fields $\{ g^i \}$ in string theory,
which arises from the summation over world-sheet topologies.
In our case, this procedure 
leads to quantum fluctuations 
in the $\sigma$-model couplings $g^i$ felt by
the propagating (low-energy) string particle modes.
The probability density ${\cal P}[g]$
in the space of $\sigma$-model theories is given 
in leading approximation by: 
\be  
{\cal P}[g] \sim {\rm exp}\left(-g^{\hat i} 
\frac{G_{{\hat i}{\hat j}}}{\Gamma}g^{\hat j} \right)
\label{prob}
\ee
where $G_{ij} \sim <V_iV_j>$ is the Zamolodchikov metric
in theory space~\cite{zam}, for vertex operators $V_i$ corresponding 
to the couplings $g^i$. Formally, the non-trivial
probability density (\ref{prob}) arises~\cite{emnquantum,lizzi} 
as a result of the requiring the cancellation 
of modular infinities against renormalization-group divergences
associated with singular configurations in the sum over
genera~\cite{emnquantum}. The width $\Gamma $ in (\ref{prob})
depends on some positive power of the string 
coupling constant $g_s$, and the precise form is 
model-dependent~\cite{lizzi,ms}. 
The quantum uncertainties $\delta g^{i}$ 
in the couplings are found by
diagonalizing the basis in theory space $g^i$, for which
knowledge of the Zamolodchikov metric is essential.  
\pr
In the presence of fluctuations in the world-sheet topology 
of the string, which were not discussed in~\cite{aemn,nature},
the shifts (\ref{shift}) associated with the
refractive index of the vacuum will fluctuate as: 
\be 
     |\delta (\Delta \alpha _i)| \sim 
|\frac{1}{2\alpha _i}c^i_{j{\hat k}}\delta g^{{\hat k}}|
\label{dispersive}
\ee
leading in turn to stochastic quantum fluctuations in
the refractive index {\it in vacuo}. This in turn
leads to a characteristic diffusive spread in the
arrival times of photons identical energies, to
which we return later. Before that, we first
compute the Zamolodchikov metric and the related
width $\Gamma$ (\ref{prob}) in a specific model
for quantum-gravitational foam based on $D$ particles.

\section{Fluctuations in $D$-particle Foam}

We compute the quantum flucuations in (\ref{prob})
in the particular case where the gravitational 
degrees of freedom $g^{\hat k}$ are the collective 
coordinates and/or momenta of 
a system of $N$ $D$ particles~\cite{dbranes,ms}. 
As discussed in the literature~\cite{kmw,emndbrane,lizzi,ms} such 
collective coordinates can be described by operators characterizing
the recoil induced by the scattering of string matter off the $D$-particle
background. The corresponding $\sigma$-model 
deformation is:
\be
Y_i^{ab}(x^0)=
\ell_s(Y_i^{ab}\ell_s\epsilon+U_i^{ab}x^0)\epsilon\Theta_\epsilon(x^0)
\label{gal}
\ee
where
the spatial coordinates of the $D$ particle 
are identified with the couplings $Y_i$ and
the $U_i$ correspond to their Galilean recoil 
velocities. These coordinates are to be regarded as
$\sigma$-model couplings $g^i$ in the sense of
the previous section. The parameter
$\epsilon\to0^+$ regulates the ambiguous value of $\Theta(s)$ at $s=0$, which
ensures that the $D$-particle system starts moving only at the
time $x^0=0$~\footnote{Throughout this paper, we consider the $\sigma$-model
`renormalized' velocity~\cite{emndbrane,lizzi,ms},
which is related to the `bare' velocity $U_{Bi}$ by 
$U_i = U_{Bi}/\epsilon $. It is $U_i$ which is an exactly-marginal
deformation
in a world-sheet renormalization group sense, and therefore corresponds to 
uniform $D$-particle motion for times  $t > 0$.}. It is
related to the world-sheet ultraviolet cutoff scale $\Lambda$ (measured in
units
of the world-sheet size $\Sigma$) by
$\epsilon^{-2}=2\ell_s^2\log\Lambda$, where
$\ell _s$ is the fundamental string length. For finite
$\epsilon$, the operators $Y_i, U_i$ each have an anomalous dimension
$-\frac12|\epsilon|^2<0$~\cite{kmw}, and thus lead to a relevant
deformation of
$S^*$. The corresponding renormalization-group 
equations are~\cite{lizzi}:
\be
dY_i^{ab}/dt=U_i^{ab}, \; dU_i^{ab}/dt=0, 
\label{rgemotion}
\ee
which are the Galilean equations of motion for the $D$ particles, if we
identify the time with the world-sheet scale: $t=\ell_s\log\Lambda$.
\pr
The natural geometry on the moduli space $\cal M$ of deformed conformal field
theories described by the above recoil operators (\ref{gal}) is given by
the following Zamolodchikov metric $G_{ab;cd}^{ij} = \langle
V_{ab}^i V_{cd}^j \rangle$~\cite{ms}:
\bea
G_{ab;cd}^{ij}&=&\frac{4g_s^2}{\ell_s^2}\left[\delta^{ij}\,I_N\otimes
I_N-\frac{g_s^2}6\left\{I_N\otimes\left(U^iU^j+U^jU^i\right)+
U^i\otimes U^j\right.\right.\nn\\&
&\biggl.\left.+\,U^j\otimes U^i+\left(U^iU^j+U^jU^i\right)\otimes
I_N\right\}\biggr]_{db;ca}+{\cal O}\left(g_s^6\right) ,
\label{Gfinal}\eea
where $I_N$ is the $N\times N$ identity matrix and we have renormalized $g_s$
to the time-independent coupling $g_s/|\epsilon|\ell_s$.
The canonical momentum $P_{ab}^i$ of
the $D$-particle system is given in the Schr\"odinger picture by
the expectation value of $-i\delta/\delta Y_i^{ab}$ evaluated in
a $\sigma$ model deformed by the operator $V_{ab}$,  
i.e., $P_{ab}^i \equiv \langle V_{ab}^i\rangle$~\cite{ms}:
\be
P_{ab}^i=\frac{8g_s^2}{\ell_s}\left[U^i-\frac{g_s^2}6
\left(U_k^2U^i+U_kU^iU^k+U^iU_k^2\right)\right]_{ba}+{\cal O}\left(g_s^6\right)
\label{canmom}\ee
which coincides with the contravariant velocity
$P_{ab}^i=\ell_sG_{ab;cd}^{ij}\dot Y_j^{cd}$ on $\cal M$. 
\pr
In agreement with general arguments~\cite{emn}, we note
that the moduli space dynamics can be derived~\cite{ms} from a Lagrangian
of the form
\be
{\cal L} = -\frac{\ell_s}2\dot Y_i^{ab}G_{ab;cd}^{ij}\dot Y_j^{cd}-{\cal 
C}
\label{lagrangian}
\ee
which coincides~\cite{ms} to leading order with the
non-Abelian
Born-Infeld effective action~\cite{tseytlin} for the target-space
$D$-particle dynamics:
\be
{\cal L}_{\rm NBI}=\frac1{\ell_s\bar g_s}\,Tr~{\rm
Sym}\,\sqrt{\det_{M,N}\left[\eta_{MN}\,I_N+\ell_s^2\bar g_s^2\,F_{MN}\right]}
\label{NBIaction}
\ee
where the trace $Tr$ is taken over $U(N)$ group indices,
\be
{\rm Sym}(M_1,\dots,M_n) \equiv \frac1{n!}\sum_{\pi\in S_n}M_{\pi_1}\cdots
M_{\pi_n}
\label{product}
\ee
is the symmetrized matrix product, and the components of
the dimensionally-reduced field-strength tensor are given by
$F_{0i}=\frac1{\ell_s^2}\dot{\bar Y}_{\!i}$ and $F_{ij}=\frac{\bar
g_s}{\ell_s^4}[\bar Y_i,\bar Y_j]$. In the Abelian reduction to the case
of a
single $D$ particle, the Lagrangian (\ref{NBIaction}) reduces to the usual
one
describing the free relativistic motion of a massive particle. The leading
order $F^2$ term in the expansion of (\ref{NBIaction}) is just the usual
Yang-Mills Lagrangian.
\pr
The formalism described in~\cite{ms} is a 
non-trivial application of the theory of Liouville string and
logarithmic operators. We note, moreover,
that the derivation of the Lagrangian (\ref{NBIaction}) was made
`off-shell', in 
other words we have compared generalized momenta in theory space with
those derived from the Born-Infeld lagrangian. The equivalence 
between the two formalisms extends beyond the 
equations of motion, which are the conformal-invariance
conditions of the $\sigma$ model. 
This is a central aspect of the recoil approach:
there are deviations from
the usual equations of motion for low-energy modes,
because conformal invariance is violated
by the recoil process~\cite{kmw,emndbrane},
and the Zamolodchikov world-sheet renormalization-group flow 
provides an off-shell treatment of this recoil problem. 
\pr
The leading contributions 
to the quantum fluctuations in $\cal M$
arise from pinched annulus diagrams 
in the summation over world-sheet genera of
the $\sigma$ model~\cite{emnquantum}. 
Symbolically, these lead to contributions in
the genus expansion of the form
\unitlength=1.00mm
\linethickness{0.4pt}
\be
\begin{picture}(70.00,10.00)
\LARGE
\put(5.00,2.00){\circle*{100.00}}
\put(8.00,2.00){\circle{3.00}}
\put(15.00,2.00){\makebox(0,0)[l]{$+$}}
\put(28.00,2.00){\circle*{100.00}}
\put(31.00,2.00){\circle{3.00}}
\put(28.00,5.00){\circle{3.00}}
\put(38.00,2.00){\makebox(0,0)[l]{$+$}}
\put(48.00,2.00){\circle{3.00}}
\put(51.00,2.00){\circle*{100.00}}
\put(54.00,2.00){\circle{3.00}}
\put(51.00,5.00){\circle{3.00}}
\put(61.00,2.00){\makebox(0,0)[l]{$+~\dots$}}
\end{picture}
\label{pinch}\ee
consisting of thin tubes 
of width $\delta\rightarrow 0$ 
(world-sheet wormholes) attached to 
the world-sheet surface $\Sigma$. The
attachment of each tube corresponds to inserting a bilocal pair
$V_{ab}^i(s)V_{cd}^j(s')$  on the boundary
$\partial\Sigma$, with interaction strength
$g_s^2$, and computing the string propagator along the thin tubes. There
are modular divergences of the form $\log\delta$,
which should be identified with
world-sheet divergences at lower genera~\cite{emn}, so we
set
\be
\log\delta=2g_s^\eta\log\Lambda=\frac1{\ell_s^2}g_s^\eta\epsilon^{-2}
\label{fs}
\ee
The exponent $\eta \neq 0$ in this 
Fischler-Susskind-like~\cite{fis} relation, 
which allows one to cancel logarithmic modular
divergences by relating the strip widths $\delta$ to the worldsheet 
ultraviolet scale $\Lambda$,
arises because the relation (\ref{fs}) is induced by the string
loop expansion. 
As we have argued in~\cite{ms}, the case $\eta\neq0$ allows 
direct comparison of our results 
with other results in the string literature, 
based on alternative approaches~\cite{liyoneya}. 
However, we do not determine $\eta$ here, 
since we only consider
string interactions between $D$ branes. 
Considering brane exchanges between the system of $D$ branes,
may enable the value of $\eta$ to be fixed (see
also below), but 
this point is not important for the present analysis. 
\pr
One effect of the dilute gas of world-sheet wormholes
is to exponentiate the bilocal operator, leading to a
change in the $\sigma$-model action~\cite{emnquantum,lizzi}. This
contribution can be cast into the form of a local action by rewriting it as a
Gaussian functional 
integral over wormhole parameters $\rho_i^{ab}$, as
described in the previous section, and we
arrive finally at~\cite{ms}:
\be
\sum_{\rm genera}Z_N[Y]\simeq\left\langle\int_{\cal
M}D\rho~e^{-\rho_i^{ab}G_{ab;cd}^{ij}\rho_j^{cd}/2|\epsilon|^2\ell_s^2
g_s^2\log\delta}~W[\partial\Sigma;Y+\rho]\right\rangle_0
\label{genusexp}\ee
We see from (\ref{genusexp}) that the effect of this resummation
over pinched
genera is to induce quantum fluctuations of the solitonic background, giving a
statistical Gaussian spread to the $D$-particle couplings. Note that the
width of
the Gaussian distribution in (\ref{genusexp}), which we identify as the
wave function of the system of $D$ particles \cite{emn}, is
time-independent,
and represents not the spread in time of a wave packet on $\cal M$, but
rather
the true quantum fluctuations of the $D$-particle coordinates.
\pr
The corresponding spatial uncertainties can be found by diagonalizing the
Zamolodchikov metric (\ref{Gfinal}), as was
done in~\cite{ms} and will not be repeated here. 
For the case of a single $D$ particle: $a=b$,  
one arrives at the uncertainties: 
\be
\left|\Delta
X^i_{aa}\right|\equiv \Delta Y^i 
=\ell_sg_s^{\eta/2}\left(1+\frac{g_s^2}{8\pi^3}\,u^2\,
\delta^{i,1}\right)+{\cal O}\left(g_s^4\right)\geq\ell_s\,g_s^{\eta/2}
\label{minimum}\ee
for the individual $D$-particle coordinates. For $\eta=0$, the minimal
length in (\ref{minimum}) 
coincides with the standard string smearing~\cite{ven}, whereas for
$\eta=\frac23$ it matches the form of $\ell_{\rm P}$ which arises from the
kinematical
properties of $D$ particles~\cite{liyoneya}. A value $\eta\neq0$ is
more
natural, because the modular divergences should be small for 
weakly-interacting strings. We
note that the uncertainty (\ref{minimum}) is 
{\it time-independent}~\cite{ms}, which is important for
experimental tests of the phenomenon, as we discuss below. 
The coordinate uncertainties for $a\ne b$ are responsible for the emergence of
a true non-commutative structure of quantum space-time, and represent the
genuine non-Abelian
characteristics of multi-$D$-particle dynamics, but we are not
concerned with such a case here.

\section{Propagation of Photons in a $D$-Particle Foam Background}

After this general discussion, we now
discuss the propagation of photons
in the background of multiple $D$ particles, taking into
account the recoil of the latter due to the scattering of the photons. 
This is immediate in the context of~\cite{ms}: 
one simply adds 
to the argument of the determinant (\ref{NBIaction}) 
an
Abelian $U(1)$ field strength $f_{MN}=\partial_M a_N - \partial _N a_M$,
where $a_M$ denotes 
the $U(1)$ electromagnetic potential of the photon fields.  
The corresponding Born-Infeld effective 
action is considered in four-dimensional space time.
We adopt the conventional view point that
photons interacting with $D$ particles may be represented by adding to the
$\sigma$-model 
action $S_\sigma$  on the (open) world sheet an electromagnetic 
potential background as a boundary term~\cite{abouel}:
\be
   S_\sigma  \ni   \frac{e}{c}\int_{\partial \Sigma} d\tau a_M (X) \partial _\tau X^M 
\label{abouel}
\ee
in a standard (Neumann) open-string notation~\footnote{The target-space 
Born-Infeld action, including photons interacting with $D$ particles, 
may {\it alternatively }
be considered as the target-space action of a three brane, i.e. a
solitonic object in string theory with four coordinates obeying Neumann
boundary conditions and the remainder Dirichlet boundary conditions~\cite{kmw}.
It is possible that the above Neumann picture can be obtained
from the Dirichlet one by a world-sheet $T$-duality transformation, but
there are problems with this transformation at a quantum 
level~\cite{otto}. We restrict ourselves here
to the Neumann string picture.}.
\pr
The non-relativistic heavy $D$-particle~\cite{ms} 
corresponds to a time-dependent background
(\ref{gal}), whilst the photon propagator
depends on the full four-dimensional space time. 
It is important to note, though,
that in our approach the
coupled $D$-particle-photon system is {\it out of equilibrium}, 
due to the recoil process and its associated distortion of the surrounding 
space time. This is reflected in the fact that the 
resulting backgrounds do not satisfy the classical equations of motion,
as was mentioned earlier in the general context of the departure from
conformal symmetry. Moreover~\cite{kanti},
the recoil curves the 
surrounding space time, since it induces - via Liouville dressing
and the identification~\cite{emn,kanti} 
of the Liouville mode with the target time $t$ -
graviton excitations for the string, which in a $\sigma$-model 
framework correspond to graviton backgrounds with non-trivial 
off-diagonal elements:
\be
    G_{0i} \sim \epsilon ^2 U_i t \Theta (t) 
\label{graviton}
\ee
where $U_i$ is the velocity of the recoiling $D$ particle.
\pr
The element (\ref{graviton}) is obviously not Lorentz covariant,
reflecting
the spontaneous breaking of this symmetry by the ground state
of the string. The splitting between the quantum-gravitational `medium' 
and the propagating particle subsystem is not possible in a 
Lorentz-invariant way, and there is no formal reason that 
this should be so. On the contrary, spontaneous violation
of Lorentz symmetry is generic in Liouville strings~\cite{aben,emn}. 
Thus we may expect the interaction of
such graviton modes with the photons to lead to a modification of 
the photon dispersion relation, analogous to 
non-Lorentz-covariant, e.g., thermal, effects 
in conventional media. A key difference between quantum-gravitational 
effects and
those in conventional media~\cite{nature} is that the 
former increase with the energy of a probe, whilst the 
latter attenuate with increasing energy. 
\pr
Gravitational interactions 
may be incorporated in the Born - Infeld lagrangian
(\ref{NBIaction}) by replacing $\eta _{MN} \rightarrow G_{MN}$ and using 
the non-covariant expression (\ref{graviton}) when calculating amplitudes.
Thus the complete off-shell Born - Infeld Lagrangian for
the interaction of photonic matter with $D$ particles is:
\be
{\cal L}_{\rm NBI}=\frac1{\ell_s\bar g_s}\,Tr~{\rm
Sym}\,\sqrt{\det_{M,N}\left[G_{MN}\,I_N+ \ell _s^2 \frac{e^2}{c}f_{MN}(a)\,I_N 
+ \ell_s^2\bar g_s^2\,F_{MN}\right]}
\label{NBIaction2}
\ee
which is the basis for our subsequent discussion.
\pr
We first re-examine 
the possible order of magnitude of the frequency-dependent
refarctive index (\ref{shift}) induced by such effects,
concentrating first on the case of Abelian 
(single $D$-particle) defects. 
The terms in the effective action (\ref{NBIaction}) 
that are relevant for the modification (\ref{shift}) 
of the photon dispersion relation arise from the
three-point function terms in (\ref{renorm}) that
involve two photon excitations and 
one induced-graviton excitation (\ref{graviton}).
The appropriate term in a derivative expansion of the Born-Infeld
action is:
\be
{\cal L}^{photons,graviton}_{NBI} 
\ni f_{MN}(a) G_{NA} f_{AM}
\sim f_{ij}(a) U_j f_{0i} \epsilon ^2 t 
\label{effact}
\ee
where latin indices are spatial, and we used (\ref{graviton}).
Terms that are similar in order of magnitude, but not in tensorial
structure, can also be obtained by combining the $f_{MN}^2$ terms  
with the determinant $\sqrt{-{\rm det}(G_{MN})}$
of the target-space metric~\footnote{It is clear that
Lorentz-invariant interactions of powers of $f^2$ 
terms with generic ${\rm Tr}(F^{2n})$ terms
in the derivative expansion of (\ref{NBIaction2}) do not affect 
the photon dispersion relation. It is only {\it non-covariant} 
terms, such as the the induced-graviton-photon 
interactions considered above, that do so.}.
An appropriate order-of-magnitude estimate of the recoil velocity 
is:
\be
U_i \sim g_s |{\overline k}|/M_P
\label{recoilvel}
\ee
reflecting energy-momentum  
conservation in the scattering process~\cite{emndbrane,lizzi,ms},
and assuming that the momentum of the recoiling $D$ particle 
is of the same order as the energy-momentum of the photon.
Working at times $t$ after the recoil that are sufficiently large,
we may assume~\cite{kmw,emndbrane,lizzi,ms} that
\be
   \epsilon ^{-2} \sim t
\label{timeapprx}
\ee
on average. 
Using now for the graviton a linear approximation 
about flat 
Minkowski space $G_{MN} \sim \eta_{MN} + h_{MN}$, which allows one to 
consider a simple Fourier momentum-space 
decomposition of the pertinent amplitudes, 
we conclude that the effective photon-graviton interactions in (\ref{effact}) 
are of order ${\cal O}(g_s E |{\overline k}|^2/M_P )$. 
Such estimates apply to the string amplitudes $c^i_{j{\hat k}}g^{{\hat k}}
$
in 
the shift (\ref{shift}) in the single-defect case, which
therefore becomes, to leading order in a low-energy  approximation:
\be
    \Delta \alpha _i ={\cal O}(-g_s \frac{E |{\overline k}|} {M_P}))
\label{shiftabel}
\ee
Using (\ref{shift2}), we see immediately that this leads to a modified
dispersion relation for the photon: 
\be
     E = |{\overline k}| + 
{\cal O}\left(-g_s\frac{E |{\overline k}|}{M_P} )\right)
\label{disp}
\ee
The refractive index is then determined from the 
photon group velocity~\cite{nature}:
\be
 v(|{\overline k}|) \equiv \frac{\partial E}{\partial |{\overline k}|} 
\simeq 1 -  {\cal O}({2g_s |{\overline k}| \over M_P}) \simeq 1 -  {\cal
O}({2g_s E \over M_P})
\label{refind}
\ee
The sign of the refractive index is determined by the fact that 
the Born-Infeld action underlying the above analysis
prevents superluminal propagation. This sign was not determined
in our previous discussion~\cite{nature}.
As discussed there,
such a variation in the velocity of light 
will cause a spread in the arrival times of pulses
of photons, according to their energy~\cite{nature}:
\be 
   \Delta t \sim {\xi L E \over M_P}
\label{figuremerit}
\ee
The possibility of observing experimentally such a shift 
using distant astrophysical sources appears conceivable,
as dicussed in~\cite{nature} and 
reviewed in the last section of this paper.

\section{Stochastic Fluctuations in the Apparent
Velocity of Light}

We now extend the above discussion to include an
estimate of the stochastic fluctuations in the 
apparent velocity of light, and hence the refractive index, due to the
summation over higher-genus world-sheet topologies.
The general theory of this summation 
in the context of Liouville string,
discussed in section 2, leads to stochastic fluctuations
in the collective coordinates of the
$D$ particles discussed in section 3~\cite{ms}:
\be
\delta Y^i \ge \ell _s g_s^{\eta/2},
\label{deltay}
\ee
The conventional Heisenberg uncertainty
relation between the coordinates
$Y^i$ and the corresponding canonical momenta $P_j$
has been shown to take
the following form for $D$ particles, after the summation 
over genera~\cite{ms}:
\be
\delta Y^i \delta P_j \ge 2 g_s^{1 + \eta/2} \delta^{i}_j 
\label{heisenberg}
\ee
Saturating the bound (\ref{deltay}) in (\ref{heisenberg}),
we obtain the following
estimate of the uncertainty in the associated 
collective canonical momenta 
of the $D$ particles:
\be
   \delta P_i \simeq \frac{2g_s}{\ell _s} 
\label{momentumvar}
\ee
We note that this estimate of the uncertainty in the collective momentum
is independent of the exponent $\eta$, which, as already
noted, would be required to take the
value $\eta = 2/3$ in order to match results on $D$
particles in conformal string theory.
\pr
In our interpretation,
the fluctuations arising from the summation over genera
lead to a statistical superposition of theories with
different values of the couplings $g^{{\hat j}}$. 
This `stochastic environment'
is characterized by the Gaussian 
form (\ref{prob}) of the associated probability distribution in the
space of $\sigma$-model backgrounds.
A photon of given energy $E$ propagating through
the fluctuating quantum-gravitational medium 
is subject to stochastic fluctuations in its velocity,
which may be obtained from Liouville interactions of the form 
(\ref{dispersive}) by endowing  
the generalized $\sigma$-model coordinates $\{ g^{\hat k} \}$ 
with the fluctuations (\ref{deltay}) found for the  
collective coordinates $Y^i$ of the $D$-particle foam.
\pr
To see how this effect indeed induces stochastic 
fluctuations in the refractive index of the photon,
one first
calculates the amplitudes $c^i_{j{\hat k}}g^{{\hat k}}$ 
appearing in (\ref{dispersive}) using
the Born-Infeld Lagrangian (\ref{NBIaction2}), 
calculated on world sheets
with disc topology. 
Using (\ref{timeapprx}), one sees that
the leading terms in a derivative expansion
are the ones given in (\ref{effact}). The summation over world-sheet 
genera leads to stochastic fluctuations 
in $U_i$ (\ref{recoilvel}),
which are given by (\ref{momentumvar}):
\be
     \delta U_i = {g_s \delta P_i \over M_P} \sim 2g_s^2 
\label{uvel}
\ee
These stochastic fluctuations in $U_i$ 
generate fluctuations in the overall 
proportionality coefficient of the Maxwell terms 
in the effective action for the photon:
\be
\delta {\cal L}^{photons,graviton}_{NBI} 
\ni f_{MN}(a) \left( \delta G_{NA}\right) f_{AM}
\sim f_{ij}(a) \left( \delta U_j \right) f_{0i} \epsilon ^2 t 
\label{effact2}
\ee
whose Fourier transform is ${\cal O} (2 g_s^2 E |{\overline k}| /M_P) $.
This does not itself affect the velocity of the photon,
but simply renormalizes the energy scale. 
\pr
However, when one proceeds
to higher orders, one picks up contributions
that lead to stochastic fluctuations in the refractive index. 
To see this, we concentrate on terms of quadratic order
in $U_i$, stemming from terms 
in a derivative expansion of the Born-Infeld lagrangian (\ref{NBIaction2}) 
that are of the generic form 
$f_{MN}^2 {\rm Tr} F_{AB}^2 $, where 
the non-Abelian field strength $F_{MN}^{ab}$ is calculated 
by considering
$Y_i^{ab}$ (\ref{gal}) formally as a `gauge potential'.
For long times of order (\ref{timeapprx}),
the only non-vanishing components of $F_{AB}$ are $F_{0i} \sim U_i$
which lead to stochastic fluctuations in the quadratic terms of order 
\be
f_{MN}^2 \left( \delta U_i^2\right) = 4 g_s^2  f_{MN}^2 U_i  
\label{stochvel}
\ee
Thus, in order to discuss 
the leading-order effects induced by the 
stochastic fluctuations of our $D$-particle foam, as
implied by the summation over world-sheet topologies,
we should consider  
corrections to the Liouville-dressed couplings 
that go beyond quadratic order
in the $\sigma$-model couplings $\{ g^{{\hat i}} \}$.
Such corrections have been studied in~\cite{dorn}, and are not
given here explicitly. It is sufficient for our purposes to 
point out that some
are proportional to four-point amplitudes, $c^i_{jkl}$
divided by `energy denominators' $\alpha_j+\alpha _k$:
$c^i_{jkl}/(\alpha_i + \alpha_j)$,
in the limit of vanishingly small central-charge deficits
$Q$ that are of interest to us, whereas others
are of the form $c^i_{jm} c^m_{kl}$ divided
by terms of order $\alpha_i^2$. The three- and
four-point amplitudes are 
computed from the Born-Infeld lagrangian (\ref{NBIaction2}) 
above~\footnote{We again reminder the reader that,
since such higher-order corrections are world-sheet renormalization-scheme
dependent~\cite{dorn}, 
this reflects the (spontaneous) breaking 
of general covariance by our foamy ground-state when the 
Liouville field is identified with target time~\cite{emn}.}.

Thus, for long times of order (\ref{timeapprx}),
and using (\ref{recoilvel}) for $U_i$,
one finds the following estimate for
the corresponding stochastic fluctuations in (\ref{shift}):  
\be
 \delta \left(\Delta \alpha _i\right) = {4 g_s^2 E^2 \over M_P}
\label{stochfluct}
\ee
Correspondingly, we find fluctuations 
in the velocity of light in the quantum-gravitational
medium of order $\delta c \sim 8g_s^2E/M_P$,
motivating the following parametrization in the
stochastic spread in photon arrival times:
\be
\left(\delta \Delta t \right) \sim {L E \over \Lambda_{QG}}
\label{error}
\ee
where $\Lambda_{QG} \sim M_P / 8 g_s^2$.
We emphasize that,
in contrast to the variation (\ref{figuremerit})
in the refractive index,
which refers to photons of different energy, 
the fluctuation (\ref{error}) 
characterizes the statistical spread in the velocity 
of photons {\it of the same energy}. 

\section{Observational tests}

The most important signatures of the refractive index
and the stochastic fluctuation in the
velocity of light that we find are
that they increase {\it linearly} with the photon
energy (frequency). This means that they can, in
principle, easily be distinguished from more
conventional medium effects, that attenuate with
increasing energy. The effects scale inversely with
some quantum-gravitational scale characteristic of
strings and $D$ branes. We are not in a position to
estimate it numerically, but we expect it to be within a few
orders of magnitude of $M_P \sim 10^{19}$ GeV.
In standard string theories one has $\eta=2/3$ and 
$g_s^2/4\pi  \sim 1/20$, but the latter may well be
modified in a more realistic theory.
In principle, one could even envisage using upper limits
on (measurements of) the rate of broadening of
a radiation spike of definite energy (frequency)
to constrain (measure) $g_s$.
\pr
We conclude this paper by mentioning some possible observational tests of
these ideas. As has been emphasized
previously~\cite{nature}, the figure of merit for constraining the
possibility
of an energy- (frequency-)dependent refractive index {\it in vacuo}
is the combination $L \times \Delta E / \Delta t$, where $L$ is the 
distance of a source, $\Delta E$ is the range of photon energies
studied, and $\Delta t$ is the observational sensitivity to
differences in light-travel times. The latter is limited by the
durations of pulses produced by the source as well as by the
resolution of the detector. 
The corresponding figure of merit for testing the new possibility
advanced in this paper, namely a stochastic spread in light-travel
times for different photons with the same energy $E$, is simply 
$L \times E / \Delta t$. 
In practice, when comparing photons of different energies, $\Delta E$ is
often dominated
by the highest photon energy $E$ that is measured, so that
$\Delta E \sim E$, and the two figures of merit are essentially
equivalent.
\pr
Astrophysical sources offer the largest figures of merit, and
the most promising that have been considered include GRBs~\cite{nature},
AGNs~\cite{biller} and pulsars~\cite{crab}. These have all been used
already to constrain the refractive-index parameter
$M_{QG}$, and offer similar sensitivities to the
stochastic-spread parameter $\Lambda_{QG}$. 
As mentioned earlier in this paper, the $D$-brane Born-Infeld analysis
indicates that higher-energy photons should be {\it retarded} relative to
lower-energy photons, rather than advanced, and their arrival times should
be more spread out. In the Table
below, we list some of the sources that have been
considered, and the sensitivities (limits) that have been
obtained. For completeness, we have also indicated the
sensitivity that might be obtainable from a detailed
analysis of the recent GRB 990123.
We see from the Table that $M_{QG}$ cannot be much smaller than the Planck
scale $M_P \sim 10^{19}$ GeV, and that some of these astrophysical sources
may already providing sensitivities to $M_{QG}, \Lambda_{QG} \sim
10^{19}$~GeV. This provides additional fundamental-physics
motivation for such $\gamma$-ray observatories as AMS~\cite{AMS} and
GLAST~\cite{GLAST}.
\pr
Other probes of the signatures of quantum gravity that
might be provided by the unorthodox photon propagation
proposed here might be possible using laboratory
experiments, for example those testing quantum optics
and searching for gravitational waves, but we do not
explore these possibilities further in this paper~\footnote{However, we
do observe that the considerations of~\cite{interfere}
concerning interferometric signatures are not
applicable in our
framework.}. However, we think that the discussion given here
demonstrates amply the possibility that 
at least some quantum-gravity ideas may be
accessible to experimental test, and need not remain 
for ever in the realm of mathematical speculation.

\begin{center}
{\bf Table: Observational Sensitivities and Limits on $M_{QG},
\Lambda_{QG}$}
\end{center}
\begin{figure}[h]
\begin{center}
\begin{tabular}{|c||c|c|c|c|}   \hline
Source & Distance & $E$ & $\Delta t$ & Sensitivity (Limit)
\\ \hline
GRB 920229~\cite{nature} & 3000 Mpc (?) & 200 keV & $10^{-2}$ s &
$10^{16}$ GeV (?) 
\\ \hline
GRB 980425~\cite{nature} & 40 Mpc & 1.8 MeV & $10^{-3}$ s (?) & $10^{16}$
GeV (?)
\\ \hline
GRB 920925c~\cite{nature} & 40 Mpc (?) & 200 TeV (?) & 200 s & $10^{19}$
GeV (?)
\\ \hline
Mrk 421~\cite{biller} & 100 Mpc & 2 TeV & 280 s & $> 4 \times 10^{16}$ GeV
\\ \hline
Crab pulsar~\cite{crab} & 2.2 kpc & 2 GeV & 0.35 ms & $> 1.8 \times
10^{15}$ GeV
\\ \hline
GRB 990123 & 5000 Mpc & 4 MeV & 1 s (?) & $3 \times 10^{14}$ GeV (?)
\\ \hline
\end{tabular} 
\end{center} 
\end{figure}
{\it The question marks in the Table indicate uncertain inputs. Hard
limits are indicated by inequality signs.}
\vspace{0.5cm}

\noindent
{\bf Acknowledgements}

We thank Phil Allport, Gianni Amelino-Camelia, Kostas Farakos, Hans
Hofer, Subir Sarkar and S.C.C. Ting for discussions and interest.
The work of D.V.N. is supported in part by D.O.E. Grant DE-FG03-95-ER-40917.

\end{document}